\documentstyle[12pt,epsf,epsfig]{article}
\def\be{\begin{equation}}
\def\ee{\end{equation}}
\def\ba{\begin{eqnarray}}
\def\ea{\end{eqnarray}}

\def\bdm{\begin{displaymath}}
\def\edm{\end{displaymath}}

\def\bq{\begin{quote}}
\def\eq{\end{quote}}

 at 10truept

\newcommand{\beq}{\begin{equation}}
\newcommand{\eeq}{\end{equation}}
\newcommand{\beqa}{\begin{eqnarray}}
\newcommand{\eeqa}{\end{eqnarray}}

\newcommand{\nn}{\nonumber}

\def\d{{\rm d}}

\def\ltap{\ \raise.3ex\hbox{$<$\kern-.75em\lower1ex\hbox{$\sim$}}\ }
\def\gtap{\ \raise.3ex\hbox{$>$\kern-.75em\lower1ex\hbox{$\sim$}}\ }
\def\gl{\ \raise.5ex\hbox{$>$}\kern-.8em\lower.5ex\hbox{$<$}\ }
\def\roughly#1{\raise.3ex\hbox{$#1$\kern-.75em\lower1ex\hbox{$\sim$}}}

\begin{document}

\thispagestyle{empty}
\begin{flushright}
arXiv:0708.1163 [astro-ph]\\
August 2007
\end{flushright}
\vspace*{.7cm}
\begin{center}
{\Large \bf How (Not) to Palatini}\\

\vspace*{1.5cm} {\large Alberto Iglesias$^{a,}$\footnote{\tt
iglesias@physics.ucdavis.edu}, Nemanja Kaloper$^{a,}$\footnote{\tt
kaloper@physics.ucdavis.edu}, Antonio Padilla$^{b,}$\footnote{\tt
antonio.padilla@nottingham.ac.uk}} \\
\vspace*{.3cm} {\large and Minjoon Park$^{a,}$\footnote{\tt
park@physics.ucdavis.edu} }\\
\vspace{.5cm} {\em $^a$Department of Physics, University of
California, Davis, CA 95616, USA}\\
\vspace{.2cm} {\em $^b$School of Physics and Astronomy, University
Park, University of
Nottingham, \\
Nottingham NG7 2RD, UK}\\

\vspace{1cm} ABSTRACT
\end{center}
We revisit the problem of defining non-minimal gravity in the
first order formalism. Specializing to scalar-tensor theories,
which may be disguised as `higher-derivative' models with the
gravitational Lagrangians that depend only on the Ricci scalar, we
show how to recast these theories as Palatini-like gravities. The
correct formulation utilizes the Lagrange multiplier method, which
preserves the canonical structure of the theory, and yields the
conventional metric scalar-tensor gravity. We explain the
discrepancies between the na\"ive Palatini and the Lagrange
multiplier approach, showing that the na\"ive Palatini approach
really swaps the theory for another. The differences disappear
only in the limit of ordinary General Relativity, where an
accidental redundancy ensures that the na\"ive Palatini works
there.  We outline the correct decoupling limits and the strong
coupling regimes. As a corollary we find that the so-called
`Modified Source Gravity' models suffer from strong coupling
problems at very low scales, and hence cannot be a realistic
approximation of our universe. We also comment on a method to
decouple the extra scalar using the chameleon mechanism.

\vfill \setcounter{page}{0} \setcounter{footnote}{0}
\newpage

\section{Introduction}

In gravity the separation of matter and geometry may be somewhat
arbitrary. Already Einstein pondered the issue of separating the
matter stress energy from geometry, especially when facing the
issue of cosmological constant.  Thus, given that we don't know
yet what is the underlying fundamental theory of the world, if one
such exists, it is amusing to consider possibilities where some
phenomena usually attributed to gravitating matter could really be
of geometric origin. This is more so when we face the problem of
explaining the dark sector of the universe, comprising over $95\%$
of the current cosmic inventory. While there are well motivated
particle physics models that can accommodate the dark matter part
of the missing mass, there really are no universally appealing
models of dark energy. It therefore seems reasonable to ask if the
observed cosmic acceleration might not really be due to some
failure of gravity to obey Einstein's General Relativity at the
largest scales, instead of the presence of some repulsive medium
filling up the world.

Among the simplest attempts to describe deviations from General
Relativity, which attracted a good deal of attention recently, are
the so-called $f(R)$ theories \cite{cacatro}-\cite{noodi}.
However, in their simplest variant, where one declares that $R$ is
the conventional Ricci scalar built out of the metric and its
derivatives, these models are nothing but general scalar-tensor
theories \cite{dickebook,wagoner} in disguise
\cite{dolgov}-\cite{nunsol}. In fact, the history of such models
is long and well known, as exemplified by a number of works
\cite{stelle}-\cite{kko} which analyzed the perturbative spectra
of higher derivative gravity. These works showed that as long as
the higher derivative terms only appear as powers of the Ricci
scalar, one can always go to a unitary gauge, decoupling the one
extra scalar degree of freedom from the helicity-2 mode in the
metric\footnote{In more complicated derivative expansions, one
will in general meet monsters: ghosts and tachyons. If such are
avoided, one may still have vector excitations \cite{stelle}.
However when Lorentz symmetry is unbroken these only couple to
matter via higher-order terms which automatically suppresses their
long range forces.}. A simple demonstration, originally due to
\cite{TT} and resurrected recently in \cite{chiba}, goes as
follows. Starting with
\be S = \int \d^4 x \sqrt{-g}\,f(R)\,, \label{fract} \ee
one can introduce an auxiliary scalar field $\varphi$ to rewrite
the gravitational action (\ref{fract}) in a simpler way, as $S =
\int \d^4 x \sqrt{-g}\,\Big( f(\varphi) + \partial_\varphi
f(\varphi)(R - \varphi)\Big)$. Then defining a new variable $\Phi
=\partial_\varphi f(\varphi)$, and inverting this equation to find
$\varphi(\Phi)$ yields
\be S = \int \d^4 x \sqrt{-g}\, \Bigg( \Phi R -  V(\Phi) \Bigg)\,,
\label{stv} \ee
with $V(\Phi) = \varphi \partial_\varphi
f(\varphi)|_{\varphi=\varphi(\Phi)} - f(\varphi)$, which indeed is
a special case of a general scalar-tensor theory \cite{wagoner}
given by $S_{\rm BD} = \int \d^4 x \sqrt{-g}\,\Big( \Phi R -
\frac{w}{\Phi}(\nabla \Phi)^2 - V(\Phi) \Big)$, with $w=0$. Thus
it appears that resorting to $f(R)$ Lagrangians really brings
nothing new or original, instead simply being a special case of
the well known and extensively studied scalar-tensor gravity.
Actually, to be a little more accurate, we should note that the
map between the $f(R)$ theory and its scalar-tensor avatar is in
fact a Legendre transform\footnote{Which seems to naturally arise
in the context of dark energy, see the relevant 
discussion in the introduction of~\cite{legendre}.}. This implies that the scalar-tensor theory
(\ref{stv}) really describes a one parameter family of $f(R)$
models, which are obtained by Legendre-inverting (\ref{stv}) and
are counted by a free integration constant.

A different twist came from some considerations of the Palatini,
or the first order, formulation of $f(R)$ models
\cite{voll1},\cite{voll}-\cite{barrow}. It was claimed that the
Palatini formulation of such models is intrinsically {\it
different} from the standard metric formulation. In fact it should
be pointed out that this apparent discrepancy between the Palatini
and the metric formulations of scalar-tensor gravities had already
been noticed in \cite{lindstrom}-\cite{querella}. It then seemed that in
the Palatini approach in some cases the extra scalar field may
altogether disappear from the spectrum of propagating modes,
leaving in its wake only an algebraic constraint.  If true, this
would have been interesting, since it might have provided a
different avenue for modifying General Relativity while respecting
Solar System tests of gravity. This issue has been discussed
extensively in \cite{voll1},\cite{voll}-\cite{fay}, whose authors
seem to have accepted the view that the same action may yield
different dynamics when formulated in terms of the metric alone or
in terms of the connexion and the metric. This seems very
puzzling, and begs the question about the nature of the
variational procedure and the canonical structure of such
formulations. On the other hand opinions on whether such
formulations are physically reasonable or not differ. In
particular, a special case where the scalar seemed to be
non-dynamical, instead of merely inducing a matter sector
constraint, has been advertised by \cite{casasito}, under the
label of `Modified Source Gravity'. These authors hoped that large
scale dynamics of such models may be reliably approximated by FRW
universes and went on to explore it as an alternative cosmological
theory. However in \cite{flana1} Flanagan has already showed that
in such theories electrons interact differently than in the usual
electrodynamics, thus conflicting with what we see in nature.
Nonetheless, the debate about such models and various hybrids
\cite{flana2,shahin,barrow}, and their physical viability, seems
to be ongoing.

In this note we revisit the Palatini formulation of nonminimal
gravities and clarify how to define such theories. In fact, the
correct first order formulation of a general gravity theory can be
obtained by paralleling the first order formulation of gauge
theories, in particular electromagnetism. This route has already
been delineated in the classic work by C. Lanczos some fifty years
ago \cite{lanczos}. We will work explicitly with scalar-tensor
theories, because this automatically covers all the $f(R)$ models
as well. The reason is that the trick explained above
\cite{TT,chiba} that relates Eqs. (\ref{fract}) and (\ref{stv})
remains perfectly applicable in any generally covariant theory
where $R$ is a scalar, independent of how it depends on the
metric, connexion et cetera. Utilizing the Lagrange multiplier
method, to safeguard the canonical structure of the theory we will
establish clearly the correspondence between the Palatini and the
metric formulation of scalar-tensor gravities, unveiling the
limitations of the commonly used na\"ive Palatini approach. Similar 
approach has been applied recently 
in $2+1$ gravity by S. Deser \cite{stanley}, with conclusions similar to 
ours\footnote{We are grateful to S. Deser for bringing \cite{stanley}
to our attention.}
will see that the discrepancies between the na\"ive Palatini and
the standard formulations of nonminimal gravity can be attributed
to the missing terms in the canonical momentum of the scalar field
variable, which shifts the scale of strong coupling of the theory.
These differences disappear in the limit of ordinary General
Relativity, where the scalar field completely decouples from the
matter sector. In the cases without scalar self-interactions, its
perturbative couplings to matter are protected by a shift symmetry
$\phi \rightarrow \phi + {\cal C}$. In this limit, there also
arises an accidental redundancy which decouples the Lagrange
multiplier field and ensures that the na\"ive Palatini and the
properly defined theory coincide as $w \rightarrow \infty$.  We will also
outline the correct strong coupling regimes and contrast them with
the decoupling limits. As a corollary we will find that the
so-called `Modified Source Gravity' models suffer from strong
coupling problems at very low scales, and not merely additional
couplings which were discussed in \cite{flana1}. Hence such models
cannot be a realistic approximation of our universe. We will
however point out one open avenue to help suppress the scalar mode,
based on the chameleon mechanism.

The paper is organized as follows. In the next section, we will
present a mechanical toy model which serves as a simple yet
complete arena to illustrate the canonical properties of scalar
tensor gravity in first and second order formulations. We will use
it to illustrate the origin of the discrepancies between the
na\"ive Palatini approach and the correct formulation as well as
to study the symmetries which emerge in the decoupling limit. In
section 3 we will show that a general scalar-tensor gravity shares
precisely the same canonical properties and has identical
accidental symmetries as our mechanical example. Section 4 is
devoted to the discussion of decoupling versus strong coupling,
and physical interpretation of the results. We will also explain
the strong coupling problem which invalidates the `Modified Source
Gravity' there, and comment on the possibility of using
chameleons. We summarize in section 5.

Before proceeding, we need to stress one more important point. In
a significant subset of the existing literature on scalar-tensor
gravities there is still an ongoing debate about which conformal
frame of the theory is `physical', implying that different frames
lead to different physical answers. This debate is nugatory.
Physical observables are frame independent, once the complete
low-energy theory of gravity and matter is correctly specified,
including the UV regulator, which may be modelled merely as a
fixed cutoff scale. Hence we will freely hop between different
conformal frames as our convenience dictates, knowing full well
that the physical conclusions do not change. We will not revel in
a more detailed justification of this fact, merely directing an
interested reader (and perhaps an occasionally doubting one) to
the classic treatise at the source of the scalar-tensor theories
\cite{dickebook}, whose clarity and lucidity we cannot hope to
surpass. We will however stipulate precisely how the variables
change in going from one frame to another.

\section{A Mechanical Toy Model}

To immediately zero in on the dynamical issues and step over the
complications with indices, we begin our discussion with an
example from particle mechanics. It faithfully represents the
general problem with first order formulation of scalar-tensor
gravity. Let us define a canonical system with an action
\be S = \int \d t \, \Bigg( \frac{w}{2Q} x\dot Q^2 - xQ \dot Y -
\frac12 x QY^2 - xJ  \Bigg) \,  , \label{mechact} \ee
where $x$ is the analogue of $\sqrt{g} g^{\mu\nu}$, $Y$ stands for
$\Gamma^\mu_{\nu\lambda}$, and $Q$ for the Brans-Dicke field
$\Phi$. The term $xJ$ symbolizes the minimal couplings to matter
sources $J$. With field redefinitions
\be z = x Q \, , ~~~~~~~~~~~~ Q = e^{\frac{q}{\sqrt{w}}} \, ,
\label{redefs} \ee
we can rewrite this action as an analogue of Brans-Dicke theory in
the Einstein frame,
\be S = \int \d t \, \Bigg( \frac{z}{2} \dot q^2 - z \dot Y -
\frac{z}2 Y^2 -  z e^{-\frac{q}{\sqrt{w}}} J  \Bigg) \,  .
\label{mechactred} \ee
Using the standard variational technique to get the equations of
motion $\frac{\dot q^2}{2} = \dot Y + \frac{Y^2}{2} +
e^{-\frac{q}{\sqrt{w}}} J$, $(z \dot q) \dot{~} =
\frac{z}{\sqrt{w}} e^{-\frac{q}{\sqrt{w}}} J$ and $Y = \frac{\dot
z}{z}$, we can eliminate $Y$ from the last equation, to finally
obtain
\be \frac{\dot q^2}{2} =( \frac{\dot z}{z}){\dot{~}}  +\frac{\dot
z^2}{2z^2} +   e^{-\frac{q}{\sqrt{w}}} J \, , ~~~~~~~~ (z \dot q)
\dot{~} = \frac{z}{\sqrt{w}} e^{-\frac{q}{\sqrt{w}}} J \,
.\label{eqstwo} \ee
These equations are the same as what one would get by varying the
action (\ref{mechactred}) {\it after} substituting $Y = \frac{\dot
z}{z}$ directly into it:
\be S = \int \d t \, \Bigg( \frac{z}{2} \dot q^2 + \frac{\dot
z^2}{2z} - z e^{-\frac{q}{\sqrt{w}}} J \Bigg) \, .
\label{grmetricth} \ee
This is analogous to the na\"ive Palatini formulation of General
Relativity with an extra scalar field, which however happens to
have non-minimal couplings to matter $J$, parameterized by the
coupling $1/\sqrt{w}$ after the transformation to the Einstein
frame.

The confusion arises if we try to recast our theory in the
original Brans-Dicke frame in the first order language. The
relations which normally follow in the metric formulation imply $Y
= \frac{\dot x}{x}$, and setting this in the action yields $S =
\int \d t \, \Big( \frac{w}{2Q} x\dot Q^2 - xQ (\frac{\dot
x}{x}){\dot{~}} - \frac{Q}{2} \frac{\dot x^2}{x} - xJ \Big)$.
Applying the field redefinitions (\ref{redefs}) {\it now}, to this
action, yields
\be S = \int \d t \, \Bigg( \Bigl(1-\frac{1}{w} \Bigr) \,
\frac{z}{2} \dot q^2 + \frac{\dot z^2}{2z} - z
e^{-\frac{q}{\sqrt{w}}} J \Bigg) \,  , \label{secondorderbd} \ee
which clearly differs from Eq. (\ref{grmetricth}) by the presence
of the piece $\propto \frac1{w}$ in the first term. This only
disappear in the limit $w \rightarrow \infty$, when the scalar
decouples from the matter sources $J$, with decoupling here being
tantamount to requiring that $q$ is bounded. This difference is
{\it precisely} the `ambiguity' which has been noticed in the
attempts to define the Palatini form of scalar-tensor gravity
\cite{lindstrom}-\cite{querella}, \cite{voll1,flana1,flana2}.

What's going on? Recall that in the na\"ive implementation of the
first order formalism in a nonminimal theory there may be
ambiguities in how to relate the connexion to the field variables.
To resolve these ambiguities, in a given theory we should enforce
the relation between the connexion and the field derivatives, or
momenta, as a constraint, using a Lagrange multiplier, as
prescribed in \cite{lanczos}. This means that our example above,
which yields (\ref{secondorderbd}) really corresponds to picking
the initial action
\be S = \int \d t \, \Bigg( \frac{w}{2Q} x\dot Q^2 - xQ \dot Y -
\frac{x Q}{2} Y^2 - xJ + \lambda \Bigl(Y - \frac{\dot x}{x}\Bigr)
\Bigg) \,  . \label{grmechactlagr} \ee
Now we can compare this to the na\"ive Palatini formulation
without the Lagrange multiplier, whose dynamics is encoded in
(\ref{grmetricth}). Applying the field redefinition
(\ref{redefs}), we can rewrite the action (\ref{grmechactlagr}) as
\be S = \int \d t \, \Bigg( \frac{z}{2}\dot q^2 - z \dot Y -
\frac{z}{2} Y^2 - z e^{-\frac{q}{\sqrt{w}}} J + \lambda \Bigl(Y -
\frac{\dot z}{z} + \frac{\dot q}{\sqrt{w}} \Bigr) \Bigg) \,  .
\label{grmechactlagq} \ee
The straightforward variation of this action shows that the
independent equations of motion reduce to (\ref{eqstwo}) only when
$w \rightarrow \infty$, corresponding to the decoupling limit of
$q$. Indeed, we have seen that the equations (\ref{eqstwo}) follow
from the substitution of $Y =  \frac{\dot z}{z}$ which arises as
the equation of motion from action (\ref{mechactred}). We could
therefore enforce this equation by a Lagrange multiplier at no
cost, which would map (\ref{mechactred}) into the form
(\ref{grmechactlagq}), but in this case {\it without} the very
last term $\propto \frac{\dot q}{\sqrt{w}}$ in the constraint.
This term on the other hand vanishes when $w \rightarrow \infty$,
rendering the two approaches coincident and simultaneously
decoupling $q$ from the matter sector $J$.

Note that in this limit the mode $q$ becomes source-free, but it
does {\it not} disappear from the equations of motion. Indeed, it
still sources the mode $z$, and therefore it `gravitates' as a
free field. This is in fact completely consistent with the
situation in scalar-tensor gravity. In the limit $w \rightarrow
\infty$ the Brans-Dicke field does not completely drop out from
the gravitational stress-energy tensor. Instead, its
matter-induced sources $\propto T^\mu{}{}_\mu$ vanish since they
are weighed by $1/\sqrt{w}$, but its own stress energy tensor
transmutes to that of a matter-free field $\phi \sim \ln \Phi$
minimally coupled to gravity, just like $q$ here. We will confirm
this below, when we move beyond our mechanical analogy. We only
note here that this provides an avenue for understanding the
weakness of $q$~-$J$ couplings in a natural way, since it is
protected by the shift symmetry $q \rightarrow q + {\cal C}$ which
arises when $w \rightarrow \infty$. Another key question which
begs one's attention following the discussion above is, if the
equations of motion which one finds following two {\it different}
approaches to defining the canonical momenta of $\dot z$ (which is
of course what the Lagrange multiplier is enforcing) are different
in general, why do they degenerate in the decoupling limit? While
this is obvious from the equations of motion, it points to yet
another emergent symmetry in the decoupling limit.

We now prove that this is precisely what happens, and illustrate
the decoupling limit en route to our result.  Let us first define
\be y = Y + \frac{ \dot q}{\sqrt{w}} \, , \label{redef1} \ee
which allows us to rewrite the action (\ref{grmechactlagr}) as
\be S = \int \d t \Bigg( z (y - \frac{\dot z}{z} ) \frac{\dot
q}{\sqrt{w}} + (z \frac{\dot q}{\sqrt{w}})\dot{~} +
(1-\frac{1}{w}) \, \frac{z}{2} \dot q^2  - z \dot y - \frac{z}{2}
y^2  - z e^{-\frac{q}{\sqrt{w}}} J + \lambda ( y - \frac{\dot
z}{z}) \Bigg) \, . \label{bigact} \ee
Clearly, in the limit $w \rightarrow \infty$ where we hold  $z, y,
q$ finite, the original variable $Q$ related to $q$ by Eq.
(\ref{redefs}) converges to a constant, $Q \rightarrow 1 + {\cal
O}(\frac{q}{\sqrt{w}})$. Further, the variable $q$ becomes
source-free, as its mixing with $J$ vanishes in this limit.
Likewise, the first two terms in (\ref{bigact}) also drop out, we
find that $y \rightarrow Y$, and the resulting theory de facto
coincides with (\ref{mechactred}) in the limit $w \rightarrow
\infty$. Conversely, for any finite value of $w$, $y$ differs from
$Y$ as prescribed in (\ref{redef1}), and the two theories are
manifestly different from each other.

Further, let us now show that in the decoupling limit the
variational equations that follow from (\ref{bigact}) become
degenerate, and the constraint equation, found by varying
(\ref{bigact}) with respect to $\lambda$ becomes redundant. This
can be seen directly from varying the full action (\ref{bigact})
and noting that the variational equations force $\lambda = 0$ in
the limit $w \rightarrow \infty$. So one may want to set $\lambda
= 0$ directly into the action in this limit, amounting to the
na\"ive Palatini approach. Yet, when this is consistent, one
expects to have a symmetry which enforces the triviality of the
constraint and decouples the Lagrange multiplier. To uncover this
symmetry, let us consider the following deformations of the
connexion variable $y$ and the Lagrange multiplier $\lambda$,
parameterized by a free parameter $a$:
\be y \rightarrow y + a \, , ~~~~~~~~~ \lambda \rightarrow \lambda
+ az \, . \label{shift} \ee
Under this shift the action (\ref{bigact}) changes to
\ba S &=&  \int \d t \Bigg(z (y - \frac{\dot z}{z} ) \frac{\dot
q}{\sqrt{w}} + (z \frac{\dot q}{\sqrt{w}})\dot{~} +
(1-\frac{1}{w}) \, \frac{z}{2} \dot q^2 - z \dot y - \frac{z}{2}
y^2
- z e^{-\frac{q}{\sqrt{w}}} J  + \lambda ( y - \frac{\dot z}{z}) \Bigg) \nonumber \\
&& + \int \d t \Bigg( a z \frac{\dot q}{\sqrt{w}} + (\lambda
+\frac{az}{2}) a \Bigg) \, , \label{bigactsh} \ea
where the first line is precisely our starting action
(\ref{bigact}), and the second line encodes the variation induced
by the shift parameter $a$. Obviously, if we now choose $az/2 = -
\lambda$, the second term in the second line of (\ref{bigactsh})
vanishes, and so if also $w \rightarrow \infty$, the action
remains invariant under this transformation of variables!

This means that in the decoupling limit the Lagrange multiplier
drops out also, since the constraint which it enforces becomes
redundant with one of the equations of motion in the theory. To
see this, we can consider one-half of the shift (\ref{shift}),
where we only change $y \rightarrow y + a$. Now (\ref{bigact})
changes to
\ba S &=&  \int \d t \Bigg(z (y - \frac{\dot z}{z} ) \frac{\dot
q}{\sqrt{w}} + (z \frac{\dot q}{\sqrt{w}})\dot{~} +
(1-\frac{1}{w}) \, \frac{z}{2} \dot q^2 - z \dot y - \frac{z}{2}
y^2
- z e^{-\frac{q}{\sqrt{w}}} J  \Bigg) \nonumber \\
&& + \int \d t \Bigg( a z \frac{\dot q}{\sqrt{w}} + (\lambda - az)
( y - \frac{\dot z}{z}) + (\lambda - \frac{az}{2}) a \Bigg) \, ,
\label{bigactsh2} \ea
where now the top line is the action (\ref{bigact}) with $\lambda
= 0$, and the second line contains the complete dependence of the
action on the Lagrange multiplier and the shift $a$. If we now
pick $a$  such that $az = \lambda$, we can decouple the Lagrange
multiplier from the connexion. Indeed, referring to the first line
of (\ref{bigactsh}) as $S_0$, we then find
\be S = S_0 + \int \d t \Bigg( \lambda \frac{\dot q}{\sqrt{w}}  +
\frac{\lambda^2}{2z} \Bigg) \, , \label{bigactsh0} \ee
where $S_0$ is totally independent of $\lambda$. Clearly when $w
\rightarrow \infty$ and $q$ is finite, $\lambda$ completely
decouples from the theory. Its dependence of the action in this
limit reduces to
\be S = S_0 + \int \d t \, \frac{\lambda^2}{2z} \, ,
\label{bigactshw0} \ee
and we can integrate it out. Alternatively, since the action is
quadratic and algebraic in $\lambda$, its variational equation in
this gauge is simply $\lambda = 0$ independently of any other
variables, and we can drop  it from the theory. The constraint
which $\lambda$ was introduced to enforce will follow from the
nontrivial variational equations that remain, as it must since it
follows in the original gauge. Therefore, we see that in the
decoupling limit, and only then, we can simply substitute the
constraint $Y  = \frac{\dot z}{z}$ in the action while maintaining
the canonical structure of the theory because of the accidental
symmetry.

\section{Beyond Mechanics: a Scalar, Gravity, and Matter}

We now turn to some generalized scalar-tensor gravity given by a
generalization of the Brans-Dicke action \cite{wagoner}
\be\label{eqn:gravact1} S_{\rm BD} = \int \d^4 x \sqrt{-g}\,\Bigg(
\Phi R - \frac{w}{\Phi}\nabla^\mu\Phi\nabla_\mu\Phi - V(\Phi) -
{\cal L}_{\rm matter}(g^{\mu\nu},\Psi) \Bigg) \, . \ee
This action, as we discussed above, includes all $f(R)$ models
after suitable field redefinitions \cite{TT,chiba,flana1}. Here we
work with the convention $R_{\mu\nu} =
\partial_{\rho}\Gamma^\rho_{\mu\nu} - \ldots $, represent the
matter fields by $\Psi$, and assume, for simplicity, that they are
minimally coupled to the Brans-Dicke frame metric $g_{\mu\nu}$. In
this, we also take all the space-time covariant derivatives which
may appear in the matter action to depend on the same connexion
coefficients $\Gamma^\mu_{\nu\lambda}$ as the one which appear in
the curvature $R$. Then, following the standard prescription for
the first order formulation \cite{lanczos}, explained in the
mechanical example of the previous section, we add to the action
(\ref{eqn:gravact1}) the Lagrange multiplier terms which enforce
Brans-Dicke frame metric compatibility\footnote{One can choose
to enforce metric compatibility in any other frame conformally related to
Brans-Dicke by a function of $\Phi$. This would yield a continuous
family of scalar-tensor theories parameterized by $w(\Phi)$.},,
\be S_{\rm LM} = \int \d^4 x \sqrt{-g}\,\Phi \lambda^\mu_{\nu\rho}
\nabla_\mu g^{\nu\rho} \,, \ee
with the Lagrange multiplier tensor fields $\lambda^\mu_{\nu\rho}$
normalized to include the prefactor $\Phi$ for later convenience.

To go to the Einstein frame, we use the field redefinitions
\be\label{eqn:cft} \tilde g_{\mu\nu} =
e^{\frac{\phi}{\sqrt{w}M_P}} g_{\mu\nu}\,, ~~~~~~~~ \Phi =
\frac{M_P^2}{2} e^{\frac{\phi}{\sqrt{w}M_P}} \,, \ee
where we are including $\sqrt{w}$ in the redefinition
(\ref{eqn:cft}) in order to facilitate taking the decoupling limit
$w\rightarrow \infty$ as in the mechanical example of the previous
section. To compare our formulas with the familiar scalar-tensor
gravities one can simply rescale the scalar field $\phi
\rightarrow \sqrt{w} \phi$ at any desired step. Since we are
working with the first order formalism, where the connexion is an
independent variable as opposed to being given by the metric
derivatives, the conformal transformation (\ref{eqn:cft}) does
{\it not} automatically induce the shift of the connexion in
passing from one frame to another. Instead, the shift, if any,
must be derived from the equations of motion generated by varying
the full action $S = S_{\rm BD}+S_{\rm LM}$. To see how, we first
rewrite it in terms of the new variables (\ref{eqn:cft}), holding
the connexion and the Ricci tensor fixed. Using $R = g^{\mu\nu}
R_{\mu\nu} = e^{\frac{\phi}{\sqrt{w}M_P}} \tilde g^{\mu\nu}
R_{\mu\nu}$, and similarly $(\nabla \phi)^2 =
e^{\frac{\phi}{\sqrt{w}M_P}} (\tilde \nabla \phi)^2$, we find
\be\label{eqn:gravact2} S = \int \d^4 x \sqrt{-\tilde
g}\,\Bigg(\frac{M_P^2}{2} \tilde g^{\mu\nu} R_{\mu\nu} -
\frac{1}{2} (\tilde \nabla \phi)^2 + \frac{M_P^2}{2}
\lambda^\mu_{\nu\rho} \Big( \frac{1}{\sqrt{w}M_P} \tilde
g^{\nu\rho} \partial_\mu \phi + \nabla_\mu \tilde g^{\nu\rho}
\Big) - \tilde V(\phi) - \tilde {\cal L}_{\rm matter} \Bigg) \,,
\ee
where $\tilde V(\phi) =  e^{-2\phi/\sqrt{w}M_P} V(\frac{M_P^2}{2}
e^{\frac{\phi}{\sqrt{w}M_P}})$ and $\tilde {\cal L}_{\rm matter} =
e^{-2\phi/\sqrt{w}M_P} {\cal L}_{\rm
matter}(e^{\phi/\sqrt{w}M_P}\tilde g^{\mu\nu},\Psi)$. Next, we
redefine the connexion $\Gamma^\mu_{\nu\rho}$ according to
$\Gamma^\mu_{\nu\rho} =  \tilde \Gamma^\mu_{\nu\rho} +
c^\mu{}_{\nu\rho}$, where
\be c^{\mu}{}_{\nu\rho} = -\frac{1}{2\sqrt{w}M_P}(
2\delta^\mu_{(\nu}\partial_{\rho)}\phi - \tilde g^{\mu\lambda}
\tilde g_{\nu\rho} \partial_\lambda\phi )\, . \label{c} \ee
Then we extract the $c^{\mu}{}_{\nu\rho}$-dependent terms out of
$\tilde g^{\mu\nu} R_{\mu\nu}$ by defining $\tilde R_{\mu\nu} =
\partial_{\rho}\tilde \Gamma^\rho_{\mu\nu} - \ldots $ and $\tilde
R = \tilde g^{\mu\nu} \tilde R_{\mu\nu}$, which allows us to
rewrite the action (\ref{eqn:gravact2}) as
\ba S &=& \int \d^4 x \sqrt{-\tilde g}\,\Bigg(\frac{M_P^2}{2}
\tilde R - \frac{1}{2}
(\tilde\nabla\phi)^2 - \tilde V(\phi) - \tilde {\cal L}_{\rm matter} \nn\ \\
&& \qquad \qquad \quad  + \,  M_P^2 \tilde g^{\mu\nu}(
\tilde\nabla_{[\rho} c^{\rho}{}_{\mu]\nu} +
c^{\rho}{}_{\lambda[\rho} c^{\lambda}{}_{\mu]\nu} ) +
\frac{M_P^2}{2} \lambda^\mu_{\nu\rho} \tilde\nabla_\mu \tilde
g^{\nu\rho}  \Bigg) \, . \label{eqn:gravact3} \ea
In this equation, we have used $\tilde \nabla$, the covariant
derivative with respect to $\tilde\Gamma$, eliminating $\nabla$
from the constraint in its favor by using Eq. (\ref{c}).

This formulation of the theory in the first order formalism
generates identical dynamics as the standard metric formulation.
One can quickly verify this by first varying the action
(\ref{eqn:gravact2}) with respect to the Lagrange multipliers,
yielding the metric compatibility condition and expressing the
connexion $\Gamma^\mu_{\nu\rho}$ in terms of the metric $\tilde
g_{\mu\nu}$ and scalar field $\phi$. Substituting this back in the
action, which is completely algebraic with respect to the
connexion apart from a boundary term, and varying the result with
respect to the metric and the scalar field $\phi$ yields precisely
the expected field equations of Brans-Dicke theory in the Einstein
frame. Indeed, let us show this for the scalar case. Eliminating
the connexion from the action, using the condition that it is the
Christoffel symbol of $e^{-\phi/\sqrt{w}M_P}\tilde g_{\mu\nu}$,
which yields
\be \tilde g^{\mu\nu} R_{\mu\nu} = \tilde R +
\frac{3}{\sqrt{w}M_P}\tilde\nabla^2\phi - \frac{3}{2w M_P^2}
(\tilde\nabla \phi)^2 \,, \ee
where $\tilde R$ is the standard Ricci scalar of $\tilde
g_{\mu\nu}$, we can rewrite (\ref{eqn:gravact2}) as
\be S = \int \d^4x \sqrt{-\tilde g}\,\Bigg( \frac{M_P^2}{2} \tilde
R - \frac{1}{2} \Big( 1+ \frac{3}{2w} \Big) (\tilde\nabla \phi)^2
- \tilde V(\phi) - \tilde{\cal L}_{\rm matter} \Bigg) \,.
\label{acteff} \ee
Varying this with respect to $\phi$ yields $( 1+\frac{3}{2w})
\tilde\nabla^2\phi - \partial_\phi \tilde V -
\frac{\delta\tilde{\cal L}_{\rm matter}}{\delta\phi} = 0$. Then we
use the stress energy tensor in the Einstein frame based on the
relation between $\tilde {\cal L}_{\rm matter}$ and ${\cal L}_{\rm
matter}$,
\be \tilde T_{\mu\nu} = -\tilde g_{\mu\nu}
e^{-\frac{2\phi}{\sqrt{w}M_P}} {\cal L}_{\rm matter} +
2 e^{-\frac{2\phi}{\sqrt{w}M_P}} \frac{\delta {\cal L}_{\rm
matter}}{\delta \tilde g^{\mu\nu}} \, , \label{strefr} \ee
and the chain rule to determine the derivatives of the Lagrangian,
which immediately yields $\frac{\delta\tilde{\cal L}_{\rm
matter}}{\delta\phi} = \frac{\tilde T }{2\sqrt{w} M_P}$. Hence the
scalar field equation is
\be \tilde\nabla^2\phi = \frac{2w}{2w+3} \frac{\partial \tilde
V}{\partial \phi}  +\frac{\sqrt{w}}{(2w+3)M_P} \tilde T \,,
\label{scfeq} \ee
precisely as in the standard metric formulation. In particular,
the limit $w \rightarrow \infty$ is the weak coupling limit, where
the theory reduces to Einstein's General Relativity, with a scalar
field with a potential $\tilde V(\phi)$, free of matter sources
and minimally coupled to gravity\footnote{We can check immediately
that the coupling of $\phi$ to gravity does not disappear but
reduces to minimal by writing the gravitational field equations,
and substituting the limit $w \rightarrow \infty$. The surviving
stress energy tensor from the scalar field $\phi$ acquires
precisely the canonical form familiar from General Relativity.}.
If in addition the potential $\tilde V$ vanishes, the theory is
invariant under a shift symmetry $\phi \rightarrow \phi + {\cal
C}$ which we noted above, and this provides an avenue for
understanding weak scalar-matter couplings from the vantage point
of naturalness. On the other hand, $w \rightarrow -\frac32$  is
the strong coupling limit, where even a tiniest self-interaction
$\partial_\phi \tilde V \ne 0$ or a source $\tilde T \ne 0$ yield
to a catastrophic response from the scalar field. If the
self-interactions and the sources are completely absent, the
scalar field does appear tame, behaving as a gravitating massless
field, which can be checked by rescaling it as $\phi \rightarrow
\phi/\sqrt{1+\frac{3}{2w}}$. Hence in this limit the scalar
behaves as a {\it poltergeist}: it needs a medium to summon it,
upon which it backreacts violently. The presence of the medium is
generic, however, which implies that the theory ceases to be under
control.

We should contrast this with what happens in the more common
na\"ive Palatini approaches \cite{lindstrom}-\cite{querella}, \cite{voll}-\cite{barrow}. 
In the na\"ive Palatini formulation,
the Lagrange multipliers and the constraints they enforce are
missing. One gets the relation between the connexion and the
metric by formally solving the algebraic equations for the
connexion which one gets from varying (\ref{eqn:gravact1}).  This
picks out Einstein frame connexion as the solution, which one can
readily verify. Hence the effective action for the metric and the
scalar field is of the same form as (\ref{acteff}), but {\it
without} the terms $\propto \frac{3}{4w} (\tilde \nabla \phi)^2$.
Clearly, this yields a different theory, which only coincides with
the standard formulation in the decoupling limit $w \rightarrow
\infty$, where the theory again reduces to General Relativity with
a minimally coupled scalar field. The strong coupling limit now is
significantly different than in the standard formulation, as can be seen from the scalar field equation,
\begin{equation}
\tilde \nabla^2 \phi=\frac{\partial \tilde V}{\partial \phi}+\frac{\tilde T}{2\sqrt{\omega} M_P}\,.
\end{equation}
In this
case, the strong coupling limit corresponds to $w \rightarrow 0$,
where the direct matter-scalar couplings explode. If the sources
vanish the field again tames up, behaving as a minimally coupled
scalar (as long as its self-interactions are  perturbative).
Again, since nonvanishing  sources are generic this outlines the
end of the validity of the theory. We shall return to more
discussion of strong coupling in the next section.

Next we explain why the difference between standard and the
na\"ive Palatini formulation disappears in the decoupling limit $w
\rightarrow \infty$. To this end, we exploit the tricks which we
have learnt in the previous section, in dealing with our simple
mechanical example, and again find an accidental symmetry in the
decoupling limit which ensures that the two approaches coincide.
We start by shifting the connexion to
\be \tilde \Gamma^\mu_{\nu\rho} \rightarrow \hat
\Gamma^\mu_{\nu\rho} =
\tilde\Gamma^\mu_{\nu\rho}+\lambda^\mu_{\nu\rho} -
\frac{2}{3}\delta^\mu_{(\nu}\lambda_{\rho)} \, , \label{conshift}
\ee
where $\lambda_\rho \equiv \lambda^\lambda_{\lambda_\rho}$ and
parenthesis in the indices denote the symmetrization of the
enclosed structures. This decouples the Lagrange multipliers
$\lambda^\mu_{\nu\rho}$ from the connexion
$\tilde\Gamma^\mu_{\nu\rho}$. Indeed, starting with the action
(\ref{eqn:gravact3}) written for the hatted connexion symbols
$\hat \Gamma^\mu_{\nu\rho}$ and substituting (\ref{conshift}), we
find
\ba\label{eqn:gravact4} S &=&  S_0 + \frac{M_P^2}{2} \int \d^4 x
\sqrt{-\tilde g}\,\Bigg( 2 \tilde g^{\mu\nu}( \tilde\nabla_{[\rho}
c^{\rho}{}_{\mu]\nu}
+ c^{\rho}{}_{\lambda[\rho} c^{\lambda}{}_{\mu]\nu} ) \nn\\
&&~~~ + \tilde g^{\mu\nu} (
\lambda^\rho_{\mu\lambda}\lambda^\lambda_{\rho\nu} - \frac{2}{3}
\lambda_\rho \lambda^\lambda_{\mu\nu} - \frac{1}{3} \lambda_\mu
\lambda_\nu ) +  \tilde g^{\mu\nu} (
\lambda^\rho_{\mu\nu}c^{\lambda}{}_{\lambda\rho} -
2\lambda^\rho_{\lambda(\mu}c^{\lambda}{}_{\nu)\rho} ) \Bigg) \,,
\ea
where $S_0 = \int \d^4 x \sqrt{-\tilde g}\,\Bigl(\frac{M_P^2}{2}
\tilde R - \frac{1}{2} (\tilde\nabla\phi)^2 - \tilde V(\phi)-
\tilde {\cal L}_{\rm matter} \Bigr) $ is the standard
Einstein-Hilbert action.

Now we can unveil the symmetry which explains why the standard
formulation and the na\"ive Palatini approach coincide in the
decoupling limit, as in the mechanical example above. Consider, in
analogy with (\ref{shift}), the transformation
\be\label{eqn:symtransf} \tilde\Gamma^\sigma_{\mu\nu}   \to
\tilde\Gamma^\sigma_{\mu\nu} + a^\sigma_{\mu\nu}  \,, ~~~~~~~~
\lambda^\sigma_{\mu\nu} \to \lambda^\sigma_{\mu\nu} +
a^\sigma_{\mu\nu}-\delta^\sigma_{(\mu}a_{\nu)}\,, \ee
where $a_\mu = a^\nu_{\mu\nu}$, under which (\ref{eqn:gravact3})
transforms into
\ba\label{eqn:gravact7} S &\to& S + \frac{M_P^2}{2} \int \d^4 x
\sqrt{-\tilde g}\, \Bigg( \tilde g^{\mu\nu}
(a_\rho c^{\rho}{}_{\mu\nu}+c_\rho a^{\rho}{}_{\mu\nu}-2a^{\rho}{}_{\mu\sigma}c^{\sigma}{}_{\rho\nu}) \nn\\
&&\qquad\qquad + \tilde g^{\mu\nu} (2\lambda^{\rho}{}_{\mu\sigma}
a^{\sigma}{}_{\rho\nu} + a^{\rho}{}_{\mu\sigma}
a^{\sigma}{}_{\rho\nu} - a_\mu a_\nu) \Bigg) \,, \ea
up to a surface term that we neglect. In the decoupling limit $w
\rightarrow \infty$, $c^\sigma{}_{\mu\nu}$ all vanish, as is clear
from (\ref{c}) and the fact that $\phi$ must remain bounded. Thus
the action will be invariant under  (\ref{eqn:symtransf}) if there
exist fields $a^\sigma_{\mu\nu}$ such that the second line of
(\ref{eqn:gravact7}) vanishes. It is straightforward to prove that
this happens for
\be\label{eqn:c3} a^\sigma{}_{\mu\nu} =
-2\lambda^\sigma{}_{\mu\nu} + \frac{10}{9}
\delta^\sigma{}_{(\mu}\lambda_{\nu)} -
\frac{2}{9}\delta^\sigma_{(\mu}\tilde g_{\nu)\rho}\tilde
g^{\alpha\beta}\lambda^\rho{}_{\alpha\beta} + \frac{4}{9}\tilde
g_{\mu\nu}\tilde g^{\alpha\beta}\lambda^\sigma_{\alpha\beta} -
\frac{2}{9}\tilde g_{\mu\nu} \tilde
g^{\sigma\alpha}\lambda_\alpha\,. \ee
Therefore, in the $w\to \infty$ limit the theory enjoys the extra
symmetry (\ref{eqn:symtransf}) with the transformation parameters
related by (\ref{eqn:c3}). This points to the fact that the
Lagrange multipliers are really irrelevant in the decoupling
limit, since their value can be altered by the action of the
symmetry transformation, without changing the action and the field
equations. Indeed, if we return to the action
(\ref{eqn:gravact4}), we see that when $w\to\infty$, $c\to 0$ it
simply becomes
\be\label{eqn:gravact5} S = S_0 + \frac{M_P^2}{2} \int \d^4 x
\sqrt{-\tilde g}\, \tilde g^{\mu\nu} (
\lambda^\rho_{\mu\lambda}\lambda^\lambda_{\rho\nu} - \frac{2}{3}
\lambda_\lambda \lambda^\lambda_{\mu\nu} - \frac{1}{3} \lambda_\mu
\lambda_\nu ) \,. \ee
This action depends on $\lambda^\mu_{\nu\lambda}$ and its
contractions only quadratically, such that its variation yields
the equations $\lambda^\mu_{\nu\lambda}=0$, allowing us to set the
Lagrange multipliers directly to zero in the action. This leaves
us precisely with the standard Einstein-Hilbert action of General
Relativity coupled to matter and a scalar field,
\be\label{eqn:gravact6} S_0 =  \int \d^4 x \sqrt{-\tilde
g}\,\Bigg( \frac{M_P^2}{2} \tilde R - \frac{1}{2} (\tilde\nabla
\phi)^2 - \tilde V(\phi) - \tilde {\cal L}_{\rm matter} \Bigg) \,.
\ee
This, of course, agrees with what we would get by taking $w\to
\infty$ limit of the na\"ive Palatini formulation without the
Lagrange multiplier. However, the agreement will only occur in
this limit, showing what went wrong with the na\"ive approach: it
swapped the theory for something else before the dynamics even
began!

\section{Strong Coupling versus Decoupling}

To continue our discussion of the aspects of strong coupling, and
contrast it with decoupling, we can rewrite the effective scalar
tensor action in the Einstein frame using the variable
\be \chi = \frac{\phi}{\sqrt{w}} \, . \label{chidef} \ee
This extracts the $w$-dependence from the matter sector, and
encodes it in the wave function renormalization $Z(w)$ of the
scalar. So after integrating out the constraints, in these
variables the action becomes
\be S = \int d^4x \sqrt{-\tilde g} \Bigg( \frac{M^2_P}{2} \tilde R
- \frac{Z(w)}{2} (\tilde \nabla \chi)^2 - \tilde V(\chi) - \tilde
{\cal L}_{matter} \Bigg) \, , \label{chiact} \ee
where now $\tilde V(\chi) =  e^{-2\chi/M_P} V(\frac{M_P^2}{2}
e^{\frac{\chi}{M_P}})$ and $\tilde {\cal L}_{\rm matter} =
e^{-2\chi/M_P} {\cal L}_{\rm matter}(e^{\chi/M_P}\tilde
g^{\mu\nu},\Psi)$, and
\be Z(w) = \cases{w + \frac32 & for the standard formulation\,,
\cr w  & for na\"ive Palatini\,. } \label{Zs} \ee
From here on we take it that all the $w$-dependence is fully coded
in $Z(w)$, for simplicity. Our conclusions can be
straightforwardly generalized to more complicated models.

With this in mind, it is now clear that the scalar sector runs
into strong coupling as $Z(w) \rightarrow 0$. Indeed the field
equation for $\chi$ which follows from (\ref{chiact}) is
\be Z(w) \tilde \nabla^2 \chi = \partial_\chi \tilde V +
\frac{\tilde T}{2M_P} \, , \label{chieqs} \ee
which can be readily verified, for example, by substituting
(\ref{chidef}) and (\ref{Zs}) into (\ref{scfeq}). Obviously, as
$Z(w) \rightarrow 0$, the field $\chi$ cannot remain bounded in
the presence of nontrivial sources or self-interactions. Things
are however much worse quantum-mechanically. From the viewpoint of
effective field theory, the action (\ref{chiact}) should be
understood as a two-derivative truncation of an infinite series,
valid only when the terms left out are sufficiently small to be
neglected (a reader may consult \cite{nima} for more discussion
and references). To check the circumstances under which this may
be hoped for, we can use canonically normalized scalar $\hat \chi
= \sqrt{Z(w)} \chi$, which implies that all of its polynomial
self-couplings $\propto g^{(n)} \chi^n$ are given by $\hat g^{(n)}
= g^{(n)}/{Z^{n/2}(w)}$, and its coupling to matter is set by $g =
\frac{1}{\sqrt{Z(w)}M_P}$. They all diverge as $Z(w) \rightarrow
0$, implying that perturbation theory, which is the only {\it a
priori} reason for the truncation of the dynamics to
(\ref{chiact}), breaks down. Thus the theory based on
(\ref{chiact}) alone is completely unreliable when $Z(w)
\rightarrow 0$. This does not preclude a possibility that a
consistent regulator exists, where (\ref{chiact}) is replaced by
something else, which  is well behaved, restoring perturbativity
of the framework. An example for this is the regularization of the
4-Fermi theory by the Standard Model. However, such a regulator
must be found before one can attempt to extract any predictions
from (\ref{chiact}) in the strong coupling regime.

The action (\ref{chiact}) also shows that our comparison of the
scalar in strong coupling to a poltergeist is apt. If we had taken
(\ref{chiact}) seriously into, and beyond, strong coupling, and
continue changing $w$ past $Z(w) = 0$, the field would have become
a ghost. This would turn on spontaneous instabilities in the
scalar sector without the need for sources. Thus, in some sense
the strong coupling regime is a welcome indicator of the impending
deterioration of the spectrum into ghost-like modes. By itself it
does not guarantee that the theory self-regulates, but merely
indicates that we should look for a suitable completion, if such
exists. However one turns this argument, it shows that the theory
(\ref{chiact}) must not be trusted when $Z(w) \rightarrow 0$ and
beyond.

To illustrate this fact, let's consider what happens if we choose
to turn the blind eye to the strong coupling warnings going up. We
could just take the action (\ref{chiact}), and the scalar field
equation (\ref{chieqs}) and declare them valid when $Z(w) = 0$.
This is exactly the idea behind the so-called `Modified Source
Gravity' proposal of \cite{casasito}.  Indeed, the motivation
behind \cite{casasito} is the following argument: take an $f(R)$
model, as in \cite{cacatro,cdtt}, pick the scales to account for
cosmic acceleration now, allow it to be mapped to a scalar-tensor
theory, for which $w=0$ \cite{chiba}, but define the dynamics
using na\"ive Palatini approach. As we have seen above, this
precisely selects (\ref{chiact}) with the second line of
(\ref{Zs}), due to the fact that the na\"ive Palatini drops a term
from the canonical momentum of the theory. Since $w=0$ as in any
$f(R)$ this means that the resulting effective theory is exactly
in the strong coupling regime $Z(0) = 0$. Now, instead of
throwing it away, \cite{casasito} take the theory (\ref{chiact})
and substitute $Z=0$ directly in this action, choosing to trust it
as such. Then the scalar field equation (\ref{chieqs}) degenerates
into the constraint
\be
\partial_\chi \tilde V + \frac{\tilde T}{2M_P} = 0 \, ,
\label{chiconst} \ee
and the field $\chi$ appears to be completely non-dynamical: a
non-propagating mode, whose only job is as a Lagrange multiplier,
to enforce (\ref{chiconst}). One could then take this equation,
combine it with the resulting field equations governing gravity
and matter, assume that some cosmological limit exists and go on
comparing to conventional cosmology.

From the viewpoint of our previous discussion of the strong
coupling limit which we reach by taking $Z \rightarrow 0$ continuosly, 
however, the condition
(\ref{chiconst}) is an extreme {\it fine tuning}. It requires
constraining the matter sector so that its dynamics
satisfies (\ref{chiconst}) with perfect precision in order to
avoid sourcing $\tilde \nabla^2 \chi$. Thus the matter fields must be
subjected to additional, abnormal dynamics, which differs from the
parent matter theory. Flanagan has already observed this in
\cite{flana1}, noting that in the na\"ive Palatini version of
$f(R)$ gravity $\chi$ is an auxiliary field that enforces the
constraint (\ref{chiconst}), and can be integrated out of the
action inducing new operators in the matter sector. The example of
\cite{flana1} involved electrons, which pick up new derivative
couplings, in conflict with the standard Maxwell theory.

However the situation is even more serious. Not only does the
matter sector pick up new interactions, but its effective
description altogether breaks down at a very low scale, much below
the cutoff of $\sim {\rm TeV}$, up to which the Standard Model
should remain valid. In fact, when the scales of the $f(R)$ theory
are picked to generate cosmic acceleration now, the matter sector
becomes strongly coupled at about the millimeter scale! This shows
that `Modified Source Gravity' is meaningless at all scales below
the millimeter, in complete conflict with what we have so far
learned about Nature.

Let us illustrate this by using a scalar matter field, e.g. the
Higgs field that controls the Standard Model masses, as a probe.
Its flat space Lagrangian, to quadratic order in fields, is ${\cal
L}_{\rm Higgs} = \frac12 (\partial {\cal H})^2 + \frac{m^2_{\cal
H}}{2} {\cal H}^2$. In this case Eq. (\ref{chiconst}) becomes
\be M_P \partial_\chi \tilde V = \frac12 e^{-\chi/M_P} (\tilde
\nabla {\cal H})^2 + e^{-2 \chi/M_P} m_{\cal H}^2 {\cal H}^2  \, .
\label{Hconst} \ee
In any generic example, where we assume that the matter theory is
perturbative so that the right-hand side of (\ref{Hconst}) is
small, we can solve for $\chi$ by Taylor expanding about the
vacuum value $\chi_*$, and determining terms order-by-order
\cite{flana1}. For the model $f(R) = \frac{M^2_P}{2} (R -
\frac{\mu^4}{R})$ studied by \cite{cacatro,cdtt}, the scalar dual
in the na\"ive Palatini approach is the theory (\ref{chiact}) with
$Z = 0$ and $\tilde V(\chi) = M^2_P \mu^2 e^{-2 \chi/M_P}
(e^{\chi/M_P} - 1)^{1/2}$ \cite{chiba,flana1}. It has a vacuum
that fits cosmic acceleration now if $\chi_* \simeq M_P$ and $\mu
\simeq H_0 \sim 10^{-33} \,  {\rm eV}$. Around the vacuum $\chi_*$
we can approximate any such potential by $\tilde V = \Lambda^4 (1+
a \frac{\chi}{M_P} + \frac{b}{2} (\frac{\chi}{M_P})^2 + \ldots )$
where $\Lambda^4 = 3 \Omega_{\Lambda} M_P^2 H_0^2$ is the dark
energy scale now, and $a$ and $b$ are ${\cal O}(1)$ numbers. The
condition (\ref{Hconst}) then yields
\be \Lambda^4 \Bigl( b \frac{\chi}{M_P} + a \Bigr) = \frac12
e^{-\chi/M_P} (\tilde \nabla {\cal H})^2 + e^{-2 \chi/M_P} m_{\cal
H}^2 {\cal H}^2 + \ldots \, . \label{rconst} \ee
Solving for $\chi$ and substituting the result in $\tilde V +
\tilde{\cal L}_{\rm Higgs}$ produces the effective action, 
$\tilde{\cal L}^{\rm(eff)}_{\rm Higgs}$, for ${\cal H}$. 
To illustrate our point, it suffices to only determine the
first few derivative terms in the expansion, which are
\be \tilde{\cal L}^{\rm(eff)}_{\rm Higgs} = \frac12 e^{-\chi_*/M_P} (\tilde \nabla
{\cal H})^2 \Bigg({\cal O}(1) + {\cal O}(1) \times e^{-\chi_*/M_P}
\frac{(\tilde \nabla {\cal H})^2}{\Lambda^4} + \ldots \Bigg) \, .
\label{effacth} \ee
Clearly, the subleading derivative terms become dominant at the
momentum scale $k_S \sim \Lambda$. Thus at momenta $k > k_S$, or
alternatively at distances $\ell < 1/k_S$ the matter sector theory
by itself ceases to be valid, demanding a new UV completion. If
$\Lambda \sim 10^{-3} \, {\rm eV}$, this means that the Standard
Model (not just the Higgs, but everything that depends on it)
becomes nonperturbative at macroscopic scales. Thus, even if we
attempt to ignore the strong coupling of the scalar $\chi$ and treat its
equation at $Z=0$ as a constraint (\ref{chiconst}), the strong coupling problem
reappears in the effective description of the matter sector,
lowering the cutoff to whatever is the $\chi$ vacuum energy scale.
If it is chosen to explain the cosmic acceleration now, the strong
coupling occurs at a very low scale, $\Lambda \sim 10^{-3} \, {\rm
eV}$. At this scale we must bring $\chi$ back into the theory, and
subsequently face its strongly coupled dynamics.  
Hence one cannot really hide from strong coupling problem in
`Modified Source Gravity'.

So far all of our dynamical analysis has been done around the
vacuum of the theory. If one applies observational bounds to it,
one then finds that models which involve gravitationally, or more
strongly, coupled, light fields that give rise to dynamical dark energy
are excluded \cite{markkteg} or worse yet, do not even make sense.
However, an avenue for hiding such a field still remains open,
albeit only narrowly. It concerns decoupling the extra degrees of
freedom with environmental effects, using the chameleon mechanism
\cite{justin}. 
We ought to emphasise that although this mechanism  
cannot rescue `Modified Source Gravity' from strong coupling problems, 
it might help a small subclass of $f(R)$ theories 
constructed using the metric formulation, as we will now explain. 
The idea is to note that an extra degree of freedom
may have an environmental contribution to the mass, which is
natural in scalar-tensor theories, because the scalar field
equation involves terms $\propto \tilde T$ as in Eq.
(\ref{scfeq}). In dense environments the scalar would appear
heavier and hence its long range forces can be Yukawa-suppressed.
In particular, in \cite{chall} it was noted that there is a
serendipitous numerical coincidence, that the density of water, in
the units of the critical density of the universe, is close to the
ratio of the Hubble length to a millimeter. Thus, if the theory
yields an environmental scalar mass linearly proportional to the
energy density of the environment, then the scalar could be
suppressed in terrestrial conditions, and within the solar system,
while simultaneously being as light as a quintessence field at
horizon scales. The linear relationship between the mass and
density is absolutely crucial. Otherwise the field which is
massive enough on Earth to pass the bounds on the deviations from
Newton's law \cite{laboratory} will be too heavy at cosmological
scales to serve as quintessence. Conversely, if it is
quintessence-light at largest scales, it will violate these bounds
\cite{chall}. On the other hand, unless the field is
quintessence-light at horizon scales, it will not yield cosmic
acceleration \cite{chall}. Thus for generic chameleon potentials
which pass the bounds of \cite{laboratory}, one must add dark
energy by hand, in a way that does not relate it to the chameleon
field\footnote{We note that a different approach was recently 
pursued by Starobinsky \cite{starobinsky}, who proposed a 
class of $f(R)$ models with a `disappearing' 
effective cosmological term, argued to arise from purely 
geometric considerations. In this model, 
there is an instability around flat space, and a channel for copious 
cosmological production of scalar modes \cite{starobinsky}.}. 
An exception to this is the logarithmic potential $\tilde V
= -\mu^4 \ln(\frac{\hat \chi}{M})$ because it yields a mass
exactly linearly proportional to $\rho_{\rm environment}$
\cite{chall}. In this case the chameleon could be dark energy, and
still remain phenomenologically viable. Since $f(R)$ models are
all dual to such scalar-tensor theories, exploring the bounds like
in \cite{chamaR} is at least reasonable,  if one really wants to
recast the chameleon quintessence as an $f(R)$ model. However, to
have a chance to 
attribute cosmic acceleration to an $f(R)$ type action, as opposed
to just adding a cosmological constant term on top of it, one
needs to pick the functional form of $f$ which reproduces the log
potential of \cite{chall}. This is straightforward in principle,
albeit mathematically contorted in practice. Indeed, the Legendre
transformation rules, which yield $V(\Phi) = \varphi
\partial_\varphi f - f$ and $\Phi = \partial_\varphi f$  combined
with $\Phi = \frac{M^2_P}{2} e^{\sqrt{\frac23}\frac{\hat
\chi}{M_P}}$ which follows from (\ref{eqn:cft})  for a canonically
normalized scalar field of the standard formulation with $w=0$ and
the condition $\tilde V =  e^{-2\sqrt{\frac23}\hat \chi/M_P}
V(\frac{M_P^2}{2} e^{\sqrt{\frac23}\frac{\hat \chi}{M_P}})$ yield the
differential equation for $f(\varphi)$,
\be - 4 \frac{\mu^4}{M_P^4} (\partial_\varphi f)^2
\ln\Bigg(\sqrt{\frac32} \frac{M_P}{M} \ln\Bigl( \frac{2}{M_P^2}
\partial_\varphi f\Bigr) \Bigg) = \varphi \partial_\varphi f - f
\, . \label{diffeqlege} \ee
Finding the solution of this equation in closed form looks rather
intimidating. One might try to select suitable limits and
seek the solutions as a series. Even so, one can see that 
the effective coupling parameter $\alpha$
of \cite{chall}, is $\alpha = \frac{1}{2\sqrt{w+3/2}}$ (aside from an irrelevant sign convention). 
Therefore when $w=0$ this 
is too large to support slow roll, which requires $\alpha < \frac{1}{4\sqrt{3}}$ 
\cite{chall}.
However solving (\ref{diffeqlege}) really produces a one-parameter
family of $f(R)$, and so one may yet carefully select the solution where
a cosmological constant contribution and the effective light chameleon
might combine to yield acceleration at the largest scales, while still modifying 
gravity and dark energy equation of state at shorter distances.

\section{Summary}

To conclude, we have clarified how to formulate scalar-tensor
gravities, and models which reduce to them, in the first order
formalism. To get correct Palatini gravities, one should use the
Lagrange multiplier method, which preserves the canonical
structure of the theory, and therefore yields the same
scalar-tensor gravity as the standard metric formulation
\cite{lanczos}. The apparent discrepancies between the na\"ive
Palatini and the standard formulation  based on Lagrange
multipliers,  encountered in \cite{lindstrom}-\cite{querella}, \cite{voll}-\cite{barrow}, arise because
the na\"ive Palatini approach really replaces the theory for
another, by changing the canonical momenta and therefore the
dynamics, as compared to the standard approach. This explains why
the same Lagrangian seemed to yield two different theories: in
fact, they have different Hamiltonians, and so really are
different from the very outset. The differences disappear only in
the decoupling limit of the scalar sector, where the theories
reduce to the ordinary General Relativity. In this case an
accidental redundancy ensures that the na\"ive Palatini works
there, as is familiar from the old lore. We have also investigated
the decoupling limits and the strong coupling regimes of the
theory. One of our side results was to establish that the
so-called `Modified Source Gravity' models suffer from strong
coupling problems at very low scales, of order $10^{-3} \, {\rm
eV}$, and hence cannot be a realistic approximation of our
universe. Finally we observed that one might use chameleon mechanism
to help decouple the extra scalar mode. However if one wants to relate the
current cosmic acceleration to the modification of 
gravity as opposed to come from a completely separate dark energy
sector, the chameleon mechanism may only work for a narrow class
of theories, defined in our Eq. (\ref{diffeqlege}), even then requiring also
a cosmological term. In sum, while
seeking for modified gravity explanations of cosmic acceleration
remains interesting, one needs to explore more dramatic ideas in
order to go beyond the effective scalar-tensor models, which are
well known and even better constrained.

\vskip1.5cm

{\bf \noindent Acknowledgements}

\smallskip

We thank Stanley Deser for bringing \cite{stanley} to our attention, with conclusions
about Palatini formulation which are very similar to ours. 
AP thanks the UC Davis HEFTI program for support and hospitality, and Damien Easson for interesting discussions.
The research of AI, NK and MP is supported in part by the DOE
Grant DE-FG03-91ER40674.  The research of NK is supported in part
by a Research Innovation Award from the Research Corporation.


\end{document}